\documentclass[letter,traditabstract]{aa} 
\usepackage{graphicx}
\usepackage{txfonts}
\begin{document}
   \title{Quasar feedback revealed by giant molecular outflows}


   \author{C. Feruglio
          \inst{1}
          \and
          R. Maiolino \inst{2}
         \and
         	E. Piconcelli  \inst{2}
	\and 
	N. Menci  \inst{2}
	\and 
	H. Aussel  \inst{1}
	\and
	A. Lamastra  \inst{2}
	\and
	F. Fiore  \inst{2} 
          }

   \institute{Laboratoire AIM, DSM/Irfu/Service d'Astrophysique, CEA  Saclay, F 91 191 Gif-sur-Yvette, France\\
              \email{chiara.feruglio@cea.fr}
         \and
            INAF- Osservatorio Astronomico di Roma, via Frascati 33, 00040 Monteporzio Catone, Italy\\
            \email{roberto.maiolino@oa-roma.inaf.it}
             }

   \date{Received June 24, 2010; accepted July 28, 2010.}

 
  \abstract{
In the standard scenario for galaxy evolution young star-forming galaxies transform into red bulge-dominated spheroids, where star formation has been quenched. To explain such a transformation, a strong negative feedback generated by accretion onto a central super-massive black hole is often invoked. The depletion of gas resulting from quasar-driven outflows should eventually stop star-formation across the host galaxy and lead the black hole to "suicide" for starvation. 
Direct observational evidence for a major quasar feedback onto the host galaxy is still missing, since outflows previously observed in quasars are generally associated with the ionized component of the gas, which only accounts for a minor fraction of the total gas content, and typically occurring in the central regions. 
We used the IRAM PdB Interferometer to observe the CO(1--0) transition in Mrk 231, the closest quasar known. Thanks to the wide band
we detect broad wings of the CO line, with velocities up to 750 km/s and spatially resolved on the kpc scale.
Such broad CO wings trace a giant molecular outflow of about 700 M$_\odot$/year, far larger than the ongoing star-formation rate ($\sim$200 M$_\odot$/year) observed in the host galaxy.
This wind will totally expel the cold gas reservoir in Mrk 231 in about 10$^7$ yrs, therefore halting the star-formation activity on the same timescale.
The inferred kinetic energy in the molecular outflow is $\sim 1.2\times 10^{44}$ erg/s, corresponding to a few percent of the AGN bolometric luminosity, which is very close to the fraction expected by models ascribing quasar feedback to highly supersonic shocks generated by radiatively accelerated nuclear winds.
Instead, the contribution by the SNe associated with the starburst fall short by orders of magnitude to account for the kinetic energy observed in the outflow. The direct observational evidence for quasar feedback reported here provides solid support to the scenarios ascribing the observed properties of local massive galaxies to quasar-induced large scale winds.}


   \maketitle
%

\section{Introduction}
In the standard scenario for galaxy evolution emerging from both observations and models young star-forming galaxies transform into red bulge-dominated spheroids, where star-formation has been quenched (Bell et al. 2004). 
Three main drivers are proposed to explain such evolution:
conditions at the time of galaxy formation,  galaxy interactions and mergers, and AGN activity.  
The ubiquitous discovery of super-massive back holes (SMBHs) at the centre of local bulges and the correlations
between their masses and bulge properties like mass, luminosity and velocity dispersion (Ferrarese \& Ford 2005 and references therein), suggest tight links between the evolution of AGN and their host galaxies. 
Indeed, massive galaxies not only live in reach environments (see e.g. Bolzonella et al. 2010, Drory et al. 2010) but are also  formed in biased regions of the density field. Furthermore, galaxy mergers are more common in rich environments, and are thought to destabilize cold gas, therefore enhancing star-formation and also funneling gas into the nuclear region. 
Such gas eventually accretes onto the nuclear SMBH, triggering an Active Galactic Nucleus (AGN). 
The gas and dust can intercept the line of sight to the nucleus, and therefore a natural expectation is that the early, powerful AGN phase is also highly obscured.  
Once SMBHs reach masses $>10^{7-8}$ M$_\odot$ the AGN power is high enough to efficiently heat the gas and expel it from the galaxy through powerful winds (Silk \& Rees 1998, Fabian et al. 1999). 
The depletion of cold gas quenches star-formation and makes the QSO nucleus ``suicide'', forcing galaxies toward the BH-spheroid mass relation observed locally. 
Such AGN negative feedback on the star formation in the host galaxy is thought to help solving two long-standing problems of galaxy evolution scenarios:  the observed small number of massive galaxies relative to the prediction of models (without AGN feedback) and their red colors, indicative of old stellar populations (Granato et al. 2004, Di Matteo et al. 2005, Menci et al. 2006, Menci et al. 2008, Bower et al. 2006).

While observational evidence for feedback on the Intra Cluster Medium has been observed in radio loud AGN (Fabian et al. 2003; McNamanara \& Nulsen 2007), direct observational evidence for AGN feedback onto the gas in the host galaxy (out of which stars form) is still missing.
In QSOs and nearby Seyfert nuclei prominent outflows are observed, but generally
only in the ionized gas component (e.g. Crenshaw et al. 2003; Turnshek 1984; Pounds et al. 2003), which accounts only
for a minor fraction of the gas mass in the host galaxy, and are generally confined in the nuclear region on pc scales or in the
photoionization cones.  
High--velocity energetic winds, whose kinetic energy is comparable to the bolometric energy of the quasar, are observed in the galactic nuclei (Reeves et al. 2009, Moe et al. 2009, Dunn et al. 2010, Bautista et al. 2010).
Morganti et al. (2010) reported evidence for AGN induced massive and fast outflows of neutral H in powerful radio galaxies, possibly driven by the AGN jets.

The bulk of the gas in QSO hosts, i.e. the molecular phase, appears little affected by the presence of the
nuclear AGN. Indeed, most studies of the molecular gas in the host galaxies of QSOs and Seyfert galaxies have found narrow CO lines
(width of a few 100 km/s), generally tracing regular rotation patterns, with no clear evidence for prominent molecular outflows
(Downes \& Solomon 1998, Wilson et al. 2008, Scoville et al. 2003), even in the most powerful quasars at high
redshift (Solomon \& Vanden Bout 2005, Omont  2007). Yet, most of the past CO observations were obtained with relatively narrow
bandwidths, which may have prevented the detection of broad wings of the CO lines possibly associated with molecular outflows. Even
worse, many CO surveys were performed with single dish, where broad CO wings may have been confused with baseline instabilities and
subtracted away along with the continuum.

In this paper we present new CO(1--0) observations of Mrk 231 obtained with the IRAM Plateau de Bure Interferometer (PdBI).
Mrk 231 is the nearest example of a quasar object and it is the most luminous Ultra-Luminous Infrared Galaxy (ULIRG) in the local Universe (Sanders et al. 1988) with
an infrared luminosity of 3.6$\times 10^{12}$ L$_\odot$ (assuming a distance of 186 Mpc). A significant
fraction ($\sim$70\%) of its bolometric luminosity is ascribed to starburst activity (Lonsdale et al. 2003).
Radio, millimeter and near-IR observations suggest that the starbursting disk is nearly face-on
(Downes \& Solomon 1998, Carilli et al. 1998, Taylor et al. 1999). 
In particular, past CO(1-0) and (2-1) IRAM PdBI observations of Mrk 231 show evidence for a regular rotation pattern and a relatively narrow profile  (Downes \& Solomon 1998), and a molecular disk (Carilli et al. 1998). 
The existence of a quasar-like nucleus in Mrk 231 has been unambiguously demonstrated by
observations carried out at different wavelengths, that have revealed the presence of a central compact radio
core plus pc-scale jets  (Ulvestad et al. 1999), broad optical emission lines (Lipari et al. 2009)  in the
nuclear spectrum, and a hard  X-ray (2-10 keV) luminosity of $10^{44}$ erg s$^{-1}$ (Braito et al. 2004).  In
addition, both optical and X-ray data have revealed that our line of sight to the active nucleus is heavily
obscured, with a measured hydrogen column as high as N$_H = 2\times 10^{24}$ cm$^{-2}$ (Braito et al.
2004).  Mrk 231 displays clear evidence of powerful ionized outflows by the multiple broad absorption lines
(BAL) systems seen all over its UV and optical spectrum. In particular, Mrk 231 is classified as a
low-ionization BAL QSOs, a very rare subclass ($\sim$10\% of the entire population) of BAL QSOs characterized by weak [OIII] emission, in which the covering factor of the absorbing outflowing material may be near unity (Boroson \& Meyers 1992). Furthermore, giant bubbles and expanding shells on kpc-scale are visible in  deep
HST imaging (Lipari et al. 2009).  Recent observations with the \emph{Herschel} Space Observatory have revealed a
molecular component of the outflow, as traced by H$_2$O and OH molecular absorption features  (Fischer et al. 2010),
but the lack of spatial information has prevented an assessment of the outflow rate.

   \begin{figure}[t!]
   \centering
  \includegraphics[scale=0.45]{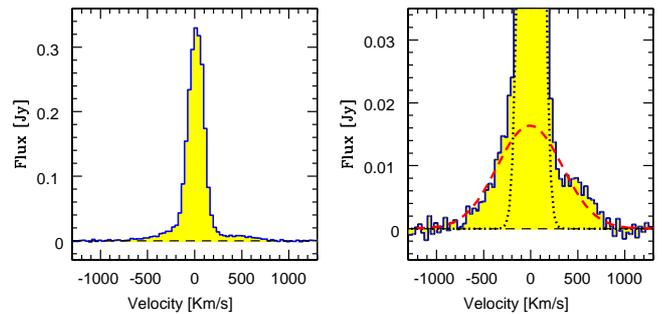}
   \caption{Continuum subtracted spectrum of the CO(1--0) transition in Mrk  231. The spectrum was extracted from a region twice the beam size (Full Width Half Maximum, FWHM), and the level of the underlying continuum emission was estimated from the region with v $>$ 800 km s$^{-1}$ and v $<$ -800 km s$^{-1}$. Left panel: full flux scale. Right panel: expanded flux scale to highlight the broad wings. The line profile has been fitted with a Gaussian narrow core (black dotted line) and a Gaussian broad component ( long-dashed line). The FWHM of the core component is 180 km s$^{-1}$ while the FWHM of the broad component is 870 km s$^{-1}$, and reaching a Full Width Zero Intensity (FWZI) of 1500 km s$^{-1}$.}
              \label{Fig1}%
    \end{figure}

\section{Data}
We have exploited the wide bandwidth offered by the PdBI to observe the CO(1-0) transition in Mrk 231.  The observations were
carried out between June and November 2009 with the PdBI, using five of the 15 m antennas of the array.
We observed the CO(1-0) rotational
transition, whose rest frequency of 115.271 GHz is redshifted to 110.607 GHz (z = 0.04217), by using using both the C and D antenna
configurations. The spectral correlator was configured to cover a band width of about 1 GHz in dual polarization. The on-source 
integration time was $\sim 20$ hours.
The data were reduced, calibrated channel by channel and analyzed by using the CLIC and MAPPING
packages of the GILDAS software. The absolute flux was calibrated on MWC349 (S(3mm) = 1.27 Jy) and 1150+497 (S(3mm) = 0.50 Jy). The absolute flux calibration error is of the order $\pm$10\%.  All maps and spectra are continuum subtracted, the continuum emission being estimated in the spectral regions with velocity v $>$ 800 km s$^{-1}$ and v $<$ -800 km s$^{-1}$.

\section{Results}
Fig. 1 shows the spectrum of the CO(1-0) emission line, dominated by a narrow component (Full Width Half Maximum $\sim$ 200 km s$^{-1}$), already detected in previous observations (Downes \& Solomon 1998, Bryant \& Scoville 1997). 
However, our new data reveal, for the first time, the presence of broad wings extending to about
$\pm$750 km s$^{-1}$, which have been missed, or possibly  confused with the underlying continuum, in previous narrower bandwidth observations.  
Both the blue and red CO(1-0) wings appear spatially resolved, as illustrated by their maps (Fig. 2).
The peak of the blue wing emission is not offset relative to the peak of the red wing, 
indicating that these wings are not due to the rotation of an inclined disk, leaving outflowing molecular gas as the only viable explanation. 
A Gaussian fit of the spatial profile of the blue and red wings (by also accounting for the beam broadening), indicates that the out-flowing medium extends over a region of about 0.6 kpc (0.7\arcsec) in radius.
To quantify the significance of the spatial extension of the high velocity, outflowing
gas, we fitted the visibilities in the {\it uv}-plane. 
We averaged the visibilities of the red and blue wings in the velocity ranges $500\div800$ km s$^{-1}$ and
$-500\div-700$ km s$^{-1}$, and we have fitted a point source, a circular gaussian and an inclined
disk model. 
The results of the {\it uv}-plane fitting are shown in Fig. 3 and summarized in Table 1. 
The upper panels of Fig. 3  show the maps of the residuals after fitting a point source model. 
The residuals of the red wing are 5$\sigma$ above the average rms of the map and those of the blue wing
3$\sigma$ above the rms.
The lower panels of Fig. 3 show the CO(1-0) wings amplitude binned in intervals
of \emph{uv} radius, covering baselines from 10 to 200 m.
The decreasing visibility amplitudes
are totally inconsistent with an unresolved source, which would instead give constant
amplitudes with radius.
The red symbols show the circular circular gaussian model fitted to the amplitudes. The fit is not perfect,
suggesting that the real spatial distribution of the red and blue CO wings
is more complex than a simple circular gaussian.
We find that the red wing is spatially resolved with a significance of $\ge 5\sigma$ while the blue wing is resolved only at 1.4$\sigma$.  The FWHM of the combined red plus blue wings is 1.42\arcsec$\pm$0.2\arcsec, corresponding to 1.2 kpc (Table 1). 
The inclined disk model gives major and minor axes of $3.2\arcsec\pm 0.48$\arcsec and $1.65\arcsec\pm0.4$\arcsec,  respectively, and a
position angle of $68\pm11$ degs. We note that
the extended blue-shifted emission is consistent with the result on Na I D by Rupke et al. (2005). 

   \begin{figure}[t!]
\centering
\begin{tabular}{cc}
\includegraphics[scale=0.26,angle=-90]{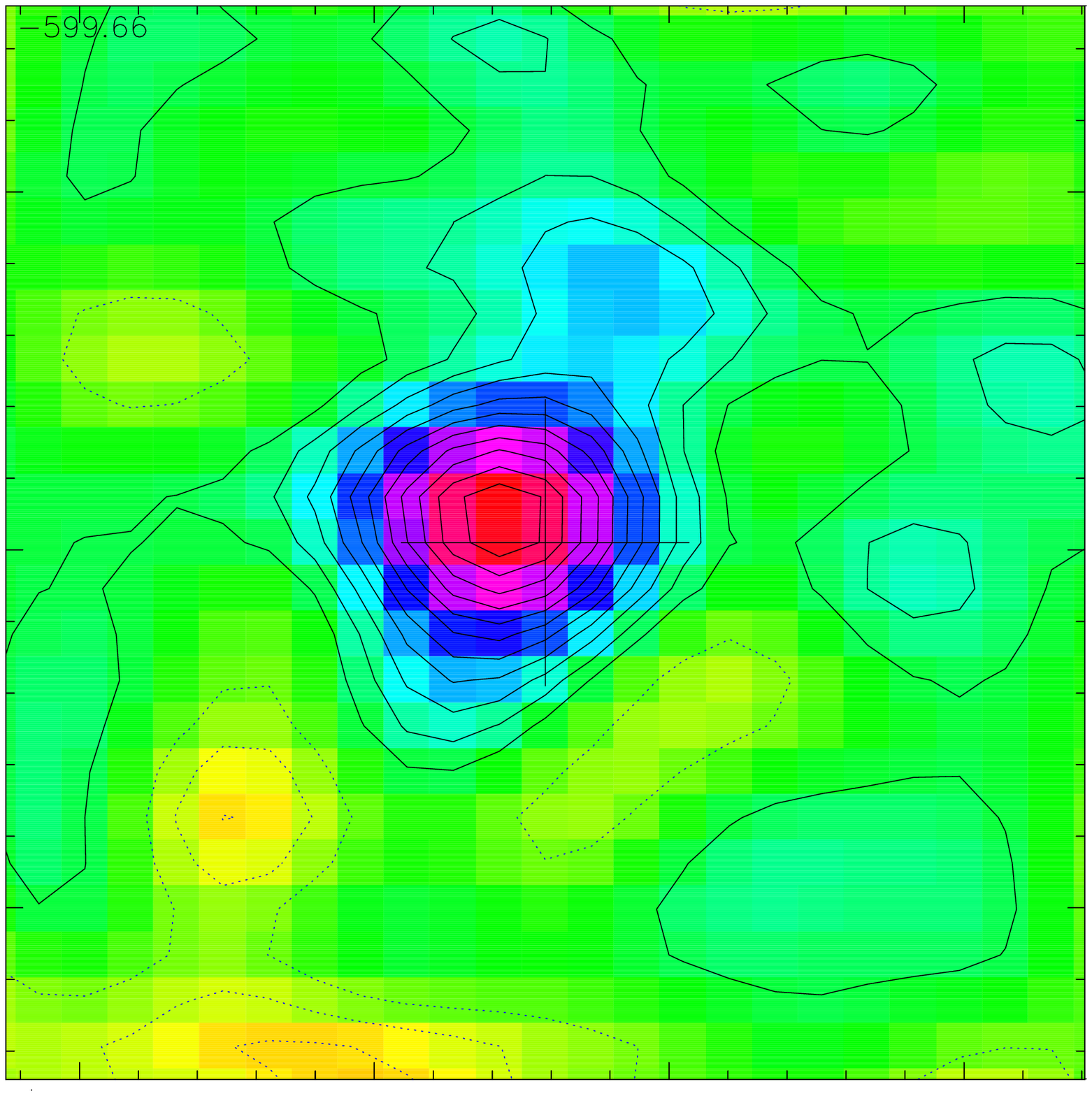}
\includegraphics[scale=0.26,angle=-90]{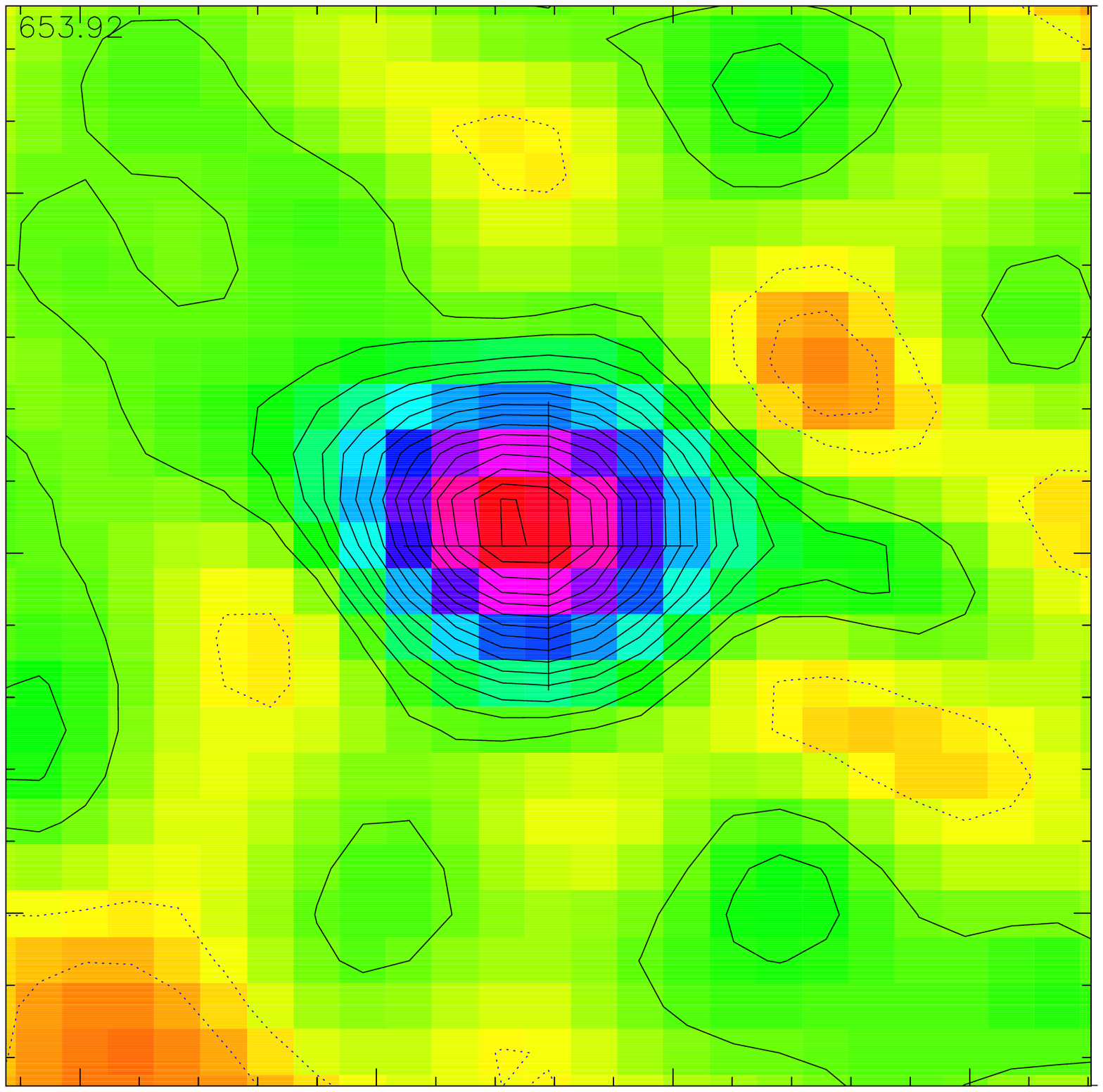}
 \end{tabular}
      \caption{CO(1-0) maps of the broad wings: the blue wing (left panel), integrated between -500 and -700 km s$^{-1}$ (right), and of the CO(1-0) red wing, integrated between 500 and 800 km s$^{-1}$ (right). The size of the maps is 15\arcsec  by 15\arcsec. 
Each contour level  corresponds to 0.2 mJy/beam.  The beam size is 2.74$\arcsec\times$ 3.4\arcsec. The cross indicates the position of the AGN.}
         \label{Fig2}
   \end{figure}

The determination of the outflow mass rate depends on the wind geometry and on the conversion factor, $\alpha$, between the CO luminosity and the molecular gas mass.  
We estimated the CO luminosity by fitting the observed line profile with a narrow plus a broad
component (see Fig. 1). The integrated CO luminosity of the broad component is L(CO)$_B = 1.16\times 10^9$ K km/s pc$^2$, about 1/10 the luminosity of the narrow component. 
We converted the CO luminosity of the broad component into molecular gas mass M(H$_2$) by assuming a
conservative conversion factor $\alpha =$ 0.5 M$_\odot$ (K km s$^{-1}$ pc$^2$)$^{-1}$, i.e. 1/10 the Galactic value . This is the lowest
conversion factor found in different locations of M 82 (a typical starburst galaxy), including its molecular outflow
(Weiss et al. 2001). We derive a mass of the outflowing molecular gas M$(H_{2 })  = 5.8\times 10^ 8$ M$_\odot$, which is consistent
with the lower limit of $7\times10^7$ M$_\odot$ inferred by Fischer et al. (2010) based on the absorption molecular lines detected by \emph{Herschel}.
By assuming that this gas is
uniformly distributed in a spherical volume of 0.6 kpc in radius, its inferred density is $\sim 25$ cm$^{-3}$. 
Since the outflow velocity is at least 700 km s$^{-1}$, the inferred mass outflow rate is $d$M$(H_2)/dt =$ 2200 M$_\odot$ yr$^{-1}$.  
If we assume density profile scaling as $r^{-2}$ for the gas distribution, the inferred outflow rate is 710 M$_\odot$ yr$^{-1}$.   
An alternative, even more conservative estimate of the
outflow rate can be derived by ignoring the Gaussian fit of the broad CO component and by using only the luminosity directly measured
from the broad wings, i.e. by integrating their flux at velocities higher than +400 km s$^{-1}$ and lower than -400 km s$^{-1}$.
The luminosity of the wings is L(CO)$ = 3.2 \times 10^8$ K km s$^{-1}$ pc$^2$  ($\sim$30\%
of the CO luminosity of the total broad component inferred through the gaussian fitting).
In this case we obtain a lower limit on the outflow rate of 600 M$_\odot$ yr$^{-1}$, for the uniform gas distribution.
By assuming a $\propto r^{-2}$ gas density profile we obtain the
most conservative lower limit is $\sim$260 M$_\odot$ yr$^{-1}$.  
We note that alternative geometries of the outflowing wind, such as shell-like or disk-like configurations, give
higher outflow mass rate, both because the inferred outflowing gas density is higher and because the de-projected outflow
velocity is higher (e.g. if the molecular outflow occurs on the galaxy disk plane). Similarly, it is easy to show
that in the case of a bipolar outflow the inferred outflow rate is the same or may be even higher
(the line of sight has to intercept
the outflowing molecular gas, since it is seen in absorption with \emph{Herschel}, implying that the true, deprojected
radius of the bipolar outflow may be equal or larger than the projected size).
In any case the inferred outflow rate is much larger than the star-formation rate in the host galaxy of ~200 M$_\odot$ yr$^{-1}$ (Taylor et al. 1999, Davies et al. 2004).
The mass loss rate being much larger than the rate at which gas is converted into stars implies a phase of rapid quenching of star formation in the regions reached by the outflow ($\sim$1 kpc scale).


 \begin{figure}[t]
\centering
\begin{tabular}{cc}
\includegraphics[scale=0.22,angle=-90]{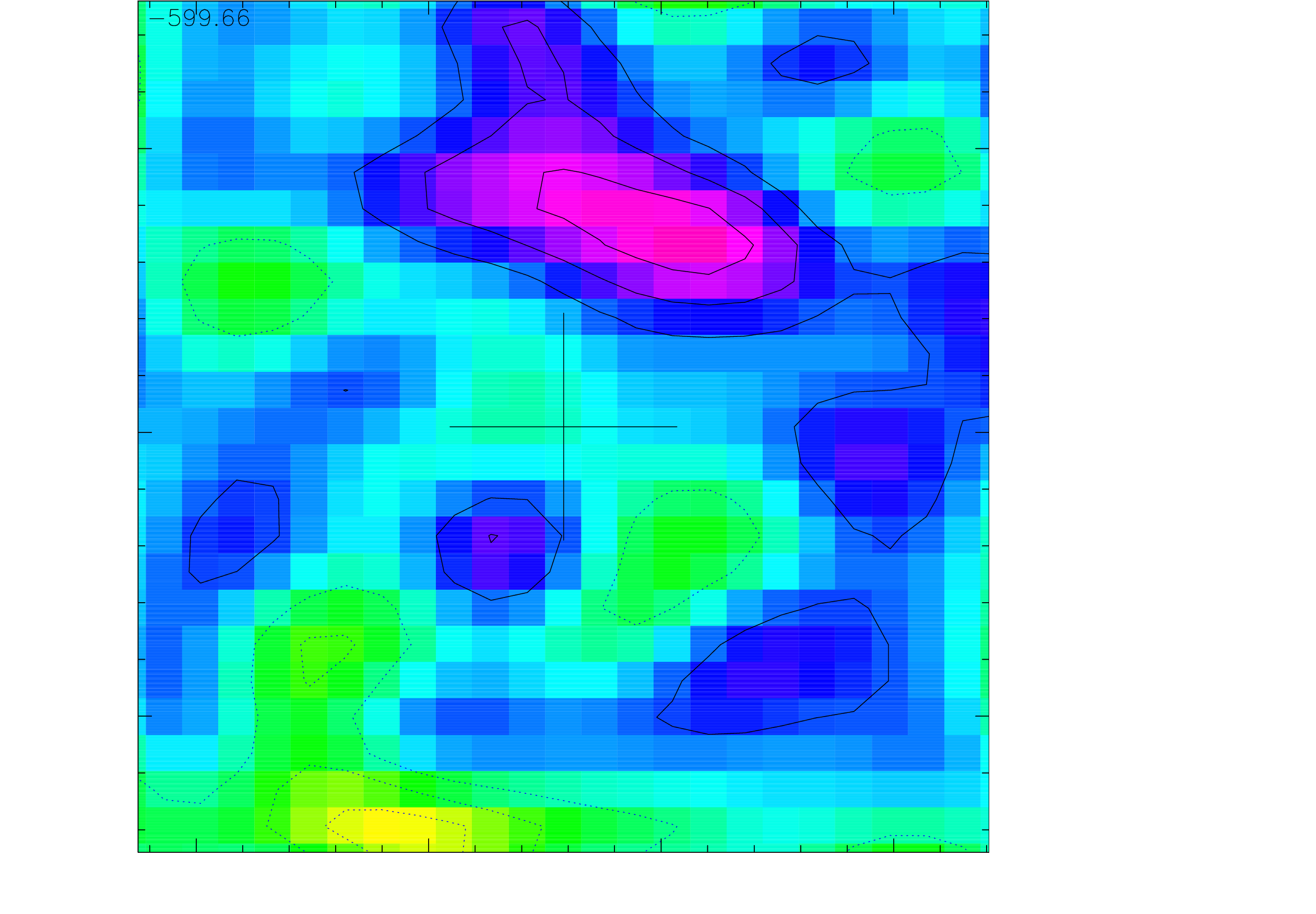}
\includegraphics[scale=0.22, angle=-90]{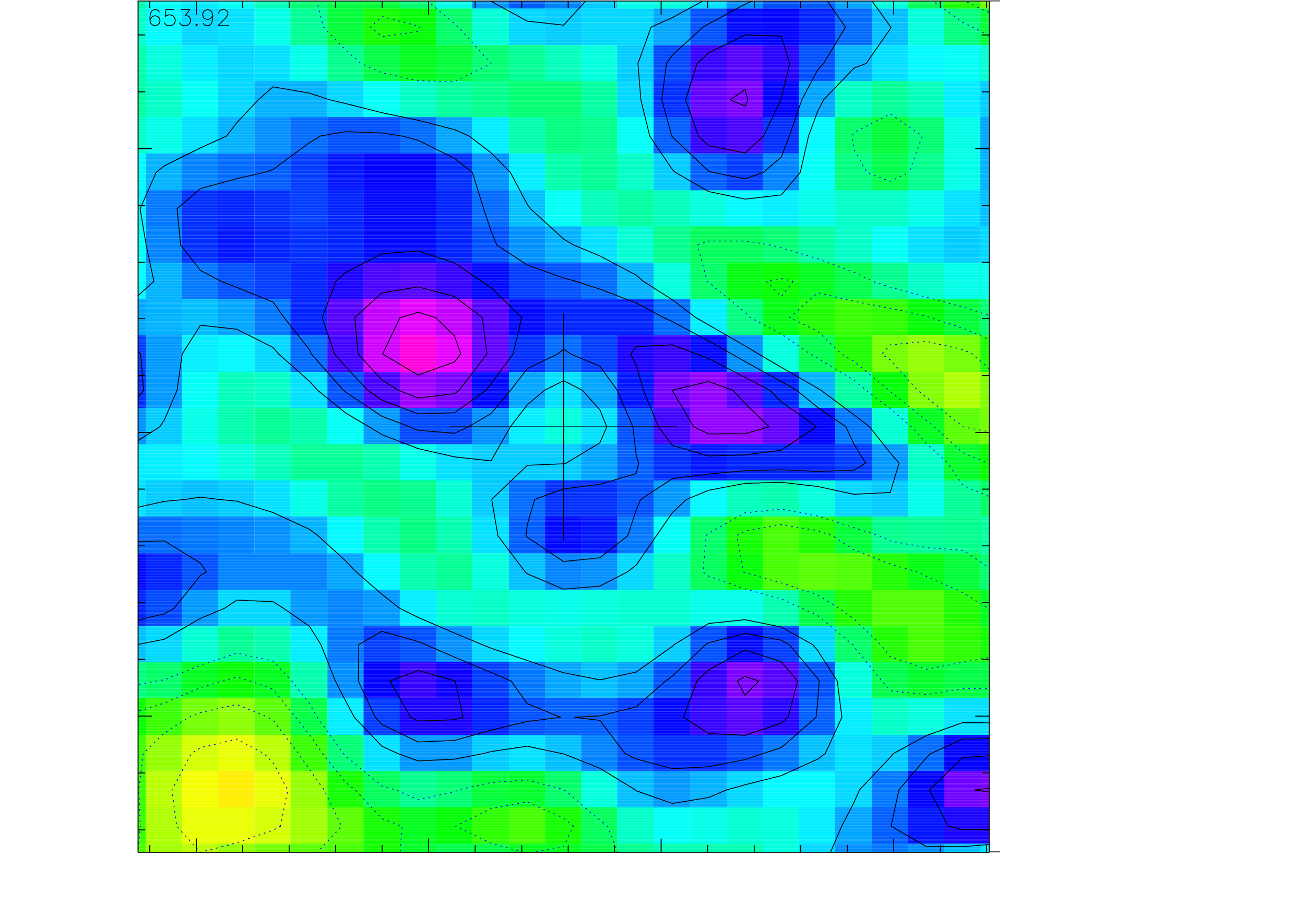}\\
\includegraphics[scale=0.22,angle=-90]{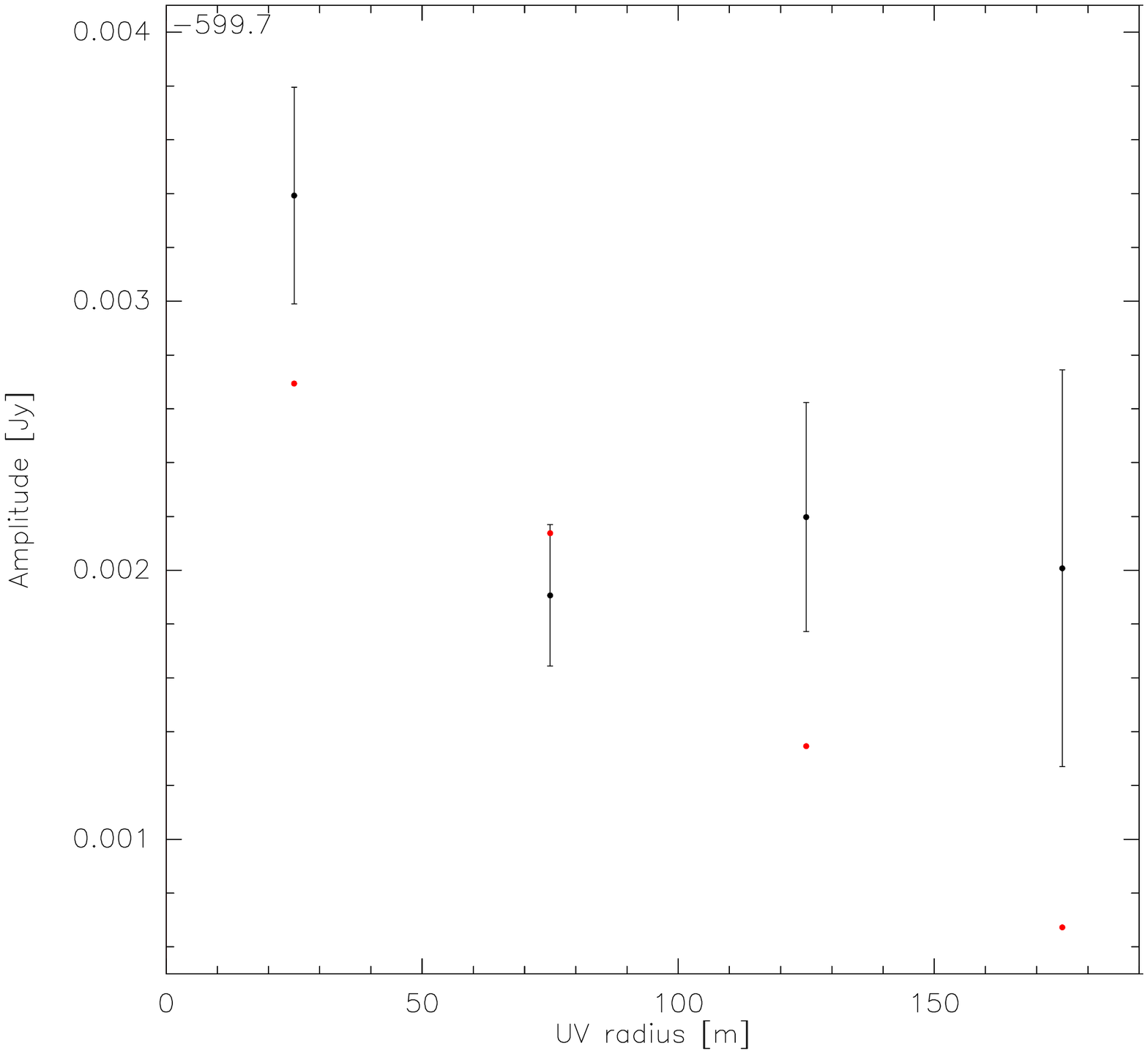}
\includegraphics[scale=0.22, angle=-90]{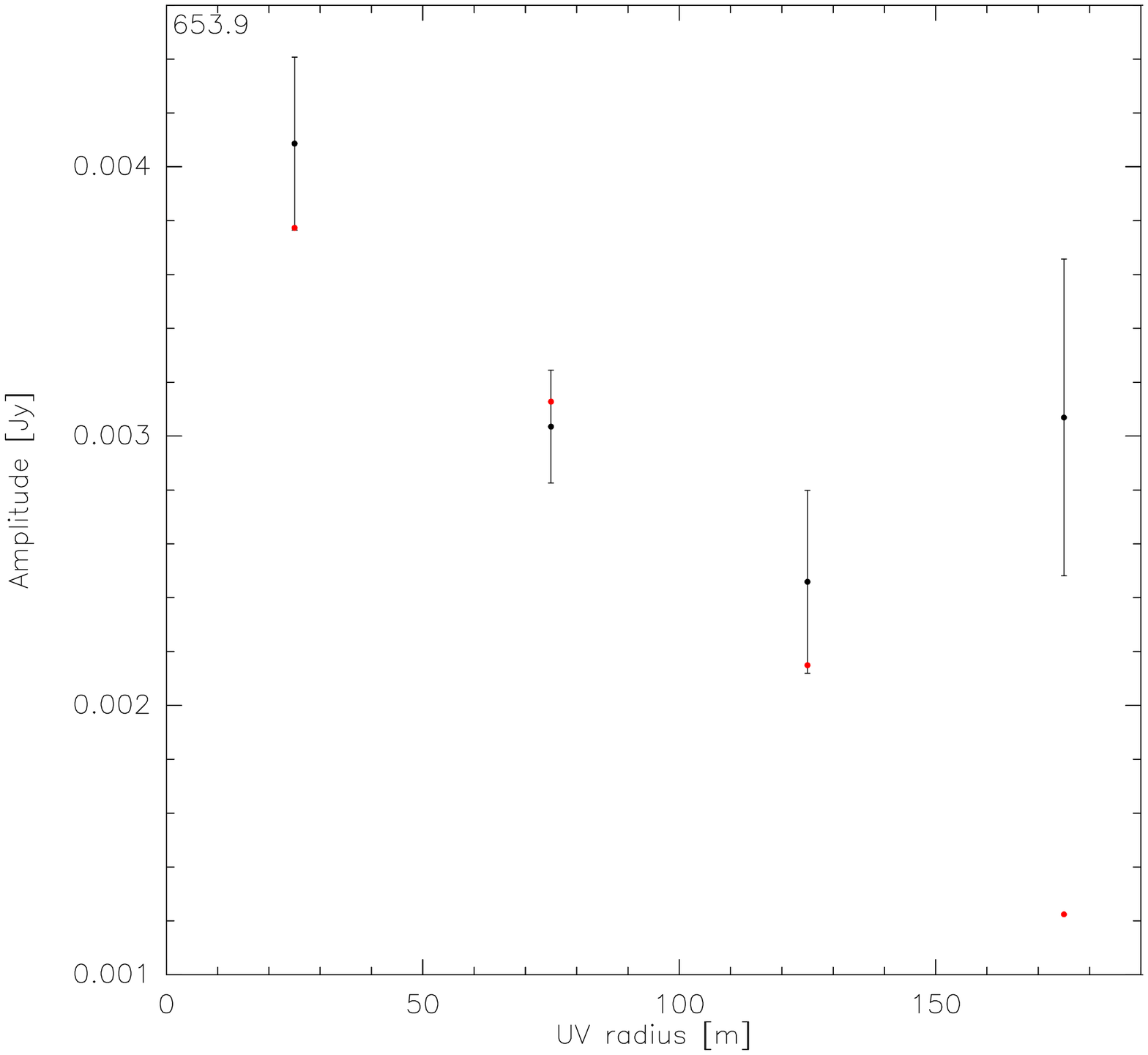}
  \end{tabular}\\
    \caption{Residual maps of a point source fit to the blue (top left panel) and red (top right panel) CO wings. The cross indicates the position of the AGN. Amplitude (in Jy) versus baseline radius for the blue (bottom left panel) and the red wings (bottom right
	panel). The red symbols show the results of a circular gaussian fit (see Tab.1). We recall that for an unresolved source the amplitude would be constant at all baselines.}\label{Fig3}
  \end{figure}

 \begin{table}[b]
 \centering
\begin{tabular}{|cccc|}
\hline
                           & velocity range & L $_{CO(1-0)}$ & FWHM CO \\
                           & (km/s)  &   K km/s pc$^2$ & (\arcsec)  \\
                           \hline
 Blue wing & $-500\div-700$ & 9.3e+07 &0.85($\pm0.61$)\\
 Red wing & $500\div800$ &  2.3e+08 &1.56($\pm0.24)$ \\
 Red+Blue w. & combined &    3.2e+08        & 1.42($\pm0.2)$\\
 \hline
 \end{tabular}
 \caption{Integration limits, luminosity and {\it uv} fit results of CO blue and red wings.}
 \end{table}

\section{Discussion and Conclusions}
The total amount of molecular gas in the galaxy disk as inferred from the integrated emission of
the narrow component, by using the CO-to-H$_2$ conversion factor $\alpha$ appropriate for ULIRGs (Solomon \& Vanden Bout 2005), is M$(H_2)\sim 10^{10}$ M$_\odot$. 
Assuming the measured outflow rate of 710 M$_\odot$ yr$^{-1}$, the total molecular gas mass will be
expelled from the host galaxy within 1.4$\times 10^{7}$ years, implying the suppression of any significant star-formation activity on the same timescale. Even with the more conservative outflow rate of 260 M$_\odot$ yr$^{-1}$ the timescale required to expel the molecular gas in the host galaxy would still be 4$\times 10^7$ years.
Such a timescale is shorter than the Salpeter time ($\sim 5\times10^7$ years).
We note that the quenching of star-formation has probably already started in the central regions.
Indeed, stellar populations younger than $5-20\times 10^6$ years are not observed within the central kpc (Lipari et al. 2009).
The total kinetic power of the out-flowing gas is 1.2$\times10^{44}$ ergs s$^{-1}$ (assuming an outflow rate of 700 km s$^{-1}$), corresponding to a few percent of the AGN bolometric luminosity, L$_{Bol}\sim  5\times10^{45}$ ergs s$^{-1}$ (Lonsdale et al. 2003) which, for a black hole mass of $\sim  6\times 10^8$ M$_\odot$ (Tacconi et al. 2002), corresponds to 6\% of the Eddington luminosity.  
Such value of the kinetic energy is very close to that expected in the case of a shock wave
produced by radiation pressure onto the interstellar medium (Lapi et al. 2005).  The corresponding energy injected into the
interstellar medium is $\Delta E \sim f \times L_{Bol} \times t_{AGN} \sim 5.5\times 10^{57}$ erg, at least 4 orders of magnitude
larger than the overall contribution of Supernovae inferred from the observed stellar mass and age of Mrk 231 (Davies et al.
2004), and even larger than the energy injected by the radio jets (Lonsdale et al. 2003).
This strongly indicates the radiation field of the QSO as the primary engine at the origin of the observed outflow.
Indeed, the observed outflow velocity of $\sim$750 km s$^{-1}$ implies a Mach number
$\mathit M \sim 30 - 50$, indicating a highly supersonic motion and, therefore, the formation of a shock front expanding in the interstellar
medium of the host galaxy.  Interestingly, this Mach number is in agreement with the expectation for the coupling between the QSO
radiation field and the gas (Lapi et al. 2005).  Indeed, the Mach number can be calculated as $M^2 \sim \Delta E/E+1$, where E$=K\times T \times M_{gas} / m_{p} = 3\times10^{54}$ erg is the thermal energy of the interstellar gas (we assumed a temperature T$=4000$ K
and an interstellar gas mass M$_{gas} \sim 10^{10}$ M$_\odot$ within the shock), and $\Delta E  = 5.5\times 10^{57}$ erg is the
overall QSO energy dumped into the interstellar medium during the time taken by the shock to reach the observed radius of 0.6 kpc.
The above computation yields  $M \sim 50$, in good agreement with our estimate based on the measured outflow speed. 
This suggests a solution to the long standing problem of the transport of energy from the nucleus to the bulk of the galaxy:
a highly supersonic shock transports outwards the energy accumulated in the centre by radiatively accelerated nuclear winds.
Extended emission and kinematic signatures in the CO line profiles similar to those observed here are predicted by hydrodynamic simulations studying the effect of AGN feedback on the molecular gas (Narayanan et al. 2006, 2008).

Due to the simultaneous presence of a strong wind, heavy X-ray absorption and of high star-formation rate, Mrk 231 has been regarded as one of the promising candidates of a QSO transiting from the obscured accretion phase, accompanied by vigorous star formation, where AGN feedback onto the host galaxy is in action, to the un-obscured phase (Page et al. 2004, Stevens et al. 2005) in the framework of the AGN-galaxy co-evolutionary sequence.  
These new observations of Mrk 231 provide one of the first direct observational evidences of QSO feedback that
is dramatically affecting the evolution of its host galaxy. 
The QSO-driven giant molecular outflow is expected to expel the disk on short
time scale, therefore halting star formation. This discovery, and in particular the inferred fraction of kinetic energy 
injected into the ISM relative to the QSO luminosity, confirms the expectations of models predicting a tight
connection between the evolution of massive galaxies and the energy released by the accreting black holes harbored in their nuclei.
The greatly improved sensitivity and expanded bandwidth of current and forthcoming millimeter interferometers will allow this kind of
studies to be extended to larger samples of QSOs. Therefore, it will be possible to directly verify whether the QSO feedback
onto the host galaxy is really ubiquitous or not.

\begin{acknowledgements}
We would like to thank Arancha Castro-Carrizo and the IRAM staff in Grenoble for helping with data reduction and calibration. We are grateful to David Elbaz, Helmut Dannerbauer and Raphael Gobat for interesting discussions. We thank the anonymous referee. We acknowledge support from grant D--SIGALE ANR--06--BLAN--01 and A--COSMOS--04--08. This work is based on observations carried out with the IRAM Plateau de Bure Interferometer. IRAM is supported by INSU/CNRS (France), MPG (Germany) and IGN (Spain).
\end{acknowledgements}

\end{document}